# Dynamic Phase Enabled Topological Mode Steering in Composite Su-Schrieffer–Heeger Waveguide Arrays

Min Tang,[1, †, *] Chi Pang,[1, †] Christian N. Saggau,[1] Haiyun Dong,[2] Ching Hua Lee,[3] Ronny Thomale,[4] Sebastian Klembt,[5] Ion Cosma Fulga,[1] Jeroen Van Den Brink,[1, 6] Yana Vaynzof,[1, 7] Oliver G. Schmidt,[8, 9, 10] Jiawei Wang,[11, **] and Libo Ma[1]

[1]Leibniz Institute for Solid State and Materials Research Dresden, Helmholtzstraße 20, Dresden 01069, Germany
[2]Key Laboratory of Photochemistry, Institute of Chemistry, Chinese Academy of Sciences, Beijing 100190, China
[3]Department of Physics, National University of Singapore, Singapore 117551, Republic of Singapore
[4]Institut für Theoretische Physik und Astrophysik, Universität Würzburg, Würzburg 97074, Germany
[5]Technische Physik, Physikalisches Institut and Wilhelm Conrad Roentgen-Research Center for Complex Material System, Universität Würzburg, D-97074 Würzburg, Germany
[6]Institute for Theoretical Physics, Technische Universität Dresden, 01069 Dresden, Germany
[7]Chair for Emerging Electronic Technologies, Technical University of Dresden, Nöthnitzer Str. 61, Dresden 01187, Germany
[8]Research Center for Materials, Architectures, and Integration of Nanomembranes (MAIN), TU Chemnitz, Chemnitz 09126, Germany
[9]Material Systems for Nanoelectronics, TU Chemnitz, Chemnitz 09107, Germany
[10]School of Science, TU Dresden, Dresden 01062, Germany
[11]School of Electronic and Information Engineering, Harbin Institute of Technology, Shenzhen, Guangdong 518055, China
†These authors contributed equally to this work.
*m.tang@ifw-dresden.de
**wangjw7@hit.edu.cn

**Abstract:** Topological boundary states localize at interfaces whenever the interface implies a change of the associated topological invariant encoded in the geometric phase. The generically present dynamic phase, however, which is energy and time dependent, has been known to be non-universal, and hence not to intertwine with any topological geometric phase. Using the example of topological zero modes in composite Su-Schrieffer-Heeger (c-SSH) waveguide arrays with a central defect, we report on the selective excitation and transition of topological boundary mode based on dynamic phase-steered interferences. Our work thus provides a new knob for the control and manipulation of topological states in composite photonic devices, indicating promising applications where topological modes and their bandwidth can be jointly controlled by the dynamic phase, geometric phase, and wavelength in on-chip topological devices.

## 1. Introduction

Topology, as a mathematical concept dealing with the invariant properties under continuous deformation, has been introduced into the field of physics to describe the characteristics of wavefunctions under a certain evolution in a parameter space. As a key topological quantity, the geometric phase (i.e., Berry phase) [1, 2] not only holds profound theoretical significance, but also leads to observable physical phenomena [3-6]. In photonics, topology has been playing a significant role in unveiling intriguing properties, such as topological robustness and one-way transmission [7-10], which has garnered increasing interest, resulting in vibrant development in applications such as lasers [11-13], filters [14-16], and beam splitters [17-19].

In addition to topology, interference is an indispensable phenomenon in wave systems, spanning from classical optics and acoustics to wavefunctions in quantum mechanics [20, 21]. So far, optical interference at topological states has been investigated to demonstrate the robustness of edge modes in various topological photonic systems [22-24], however, there has been limited exploration of interference between topological edge modes in composite photonic structures, which has the potential to open up new avenues for the manipulation of topological states in carefully designed photonic devices.

In the present work, we investigated the excitation and transition between bulk states and non-trivial topological states based on dynamic-phase-steered interference of non-trivial topological modes in a composite c-SSH waveguide array. The c-SSH is a composite system in which two SSH waveguide arrays are symmetrically distributed at the two sides of a central waveguide. Two input waveguides on the left and right sides of the central waveguide, capable of controlling the dynamic phase difference of input light therein, are used to couple laser light into the c-SSH system for the excitation of topological modes. This strategy provides a new approach for the control and utilization of topological states in composite photonic devices, offering valuable insights necessary for the realization of future applications for regulating light flow in on-chip topological systems.

## 2. Results and Discussion

Figure 1(a) schematically shows a c-SSH waveguide array system in its cross-sectional view, where two SSH waveguide arrays are coupled by a central waveguide. This composite system can be viewed as a slightly deformed SSH waveguide array model with a kinked interface [25-27], which can suppress trivial edge states, especially Tamm states [28-30]. The light propagation behavior within the c-SSH waveguide array can be described by the coupled mode theory (CMT) [31] in the tight-binding approximation:

$$i\frac{\partial a_0(z)}{\partial z} + \beta_0 a_0(z) + K_c\big(a_{+1}(z) + a_{-1}(z)\big) = 0 \quad (1)$$

$$i\frac{\partial a_j(z)}{\partial z} + \beta_0 a_j(z) + K_c a_0(z) + K_a a_{j+sign(j)}(z) = 0$$
$$(j = \pm 1) \quad (2)$$

$$i\frac{\partial a_j(z)}{\partial z} + \beta_0 a_j(z) + K_b a_{j+sign(j)}(z) + K_a a_{j-sign(j)}(z) = 0$$
$$(\text{for mod}(|j|,2) = 0) \quad (3)$$

$$i\frac{\partial a_j(z)}{\partial z} + \beta_0 a_j(z) + K_a a_{j+sign(j)}(z) + K_b a_{j-sign(j)}(z) = 0$$
$$(\text{for mod}(|j|,2) = 1), \quad (4)$$

where $j$ enumerates the relative position of waveguides ($j = 0, \pm 1, …, \pm N$) to the central waveguide ($j = 0$, see Fig 1(a)), and $a_j(z)$ denotes the complex amplitude evolving in $z$ direction which is an analog to time. $K_c$ denotes the coupling strength between the central waveguide and its neighboring waveguides, while $K_a$ and $K_b$ denote the alternating coupling strengths in the left and right waveguide arrays. All component waveguides have identical geometrical parameters with the same propagation constant $\beta_0$. As shown in Fig. 1(a), the central waveguide is marked by a red dashed box which couples with SSH waveguide arrays at each side.

The waveguides of $j = 1, …, N$ and $j = -1, …, -N$ form two SSH waveguide arrays at each side, constituting a one-dimensional topological system. The bulk Hamiltonian of an infinite SSH waveguide array takes the form [32, 33]

$$H(k) = \begin{pmatrix} 0 & \rho*(k) \\ \rho(k) & 0 \end{pmatrix}, \quad (5)$$

where k is the wavevector in the direction perpendicular to the waveguide axis $z$ and $\rho(k) = K_a + K_b e^{ik}$. As mentioned above, $K_a$ and $K_b$ are periodically alternating coupling strengths in the SSH lattices. If $K_a > K_b$, changing $k$ from 0 to $2\pi$ does not result in encircling of $\rho(k)$ around the origin point in the Brillouin zone, corresponding to the trivial phase. In contrast, the system enters the non-trivial phase regime for the case of $K_a < K_b$. The winding number as well as the Zak phase in the Brillouin zone can also be calculated via $\oint dk \langle u_k | \partial_k u_k \rangle$ [34]. In our designed c-SSH waveguide array, $K_a < K_b$ is always satisfied as the condition of $d_a > d_b$ is maintained, where $d_a$ and $d_b$ are the alternating waveguide gaps in the lattice (see Fig. 1(a)). The topological edge states were numerically simulated for the left- and right-SSH waveguide arrays, as shown in Fig. 1(b). The antisymmetric topological zero mode (TZM) [35] will be generated in the c-SSH which can be regard as a combination of the topological edge states (TESs) satisfying specified phase relationships.

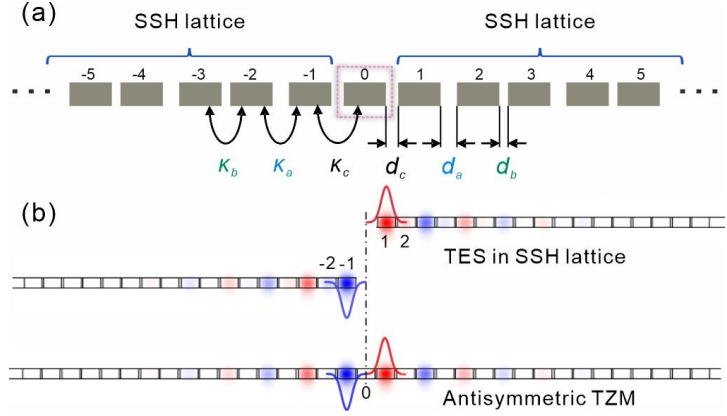

Fig. 1. (a) Schematic diagram of a composite waveguide array (cross-section) with a central defect ($j = 0$). The waveguide gaps and corresponding coupling strengths are labeled by $d_c/K_c$, $d_a/K_a$ and $d_b/K_b$, respectively. (b) Numerical simulations show TESs in the SSH waveguide lattices and an antisymmetric TZM in the composite SSH lattice. The solid lines depicting the field distribution illustrate the dynamic phase relationship between the mode fields in boundary waveguides ($j = \pm 1$). In the simulation, the width and height of the waveguide are 1.5 μm and 220 nm, respectively. The gap ($d_c$) between the central waveguide and the adjacent waveguides is 140 nm. $d_a$ and $d_b$ are 180 and 120 nm, respectively.

The TZM can also be revealed through the calculated spectrum in a c-SSH lattice with open boundaries. The Hamiltonian is derived from equations (1)-(4) under the condition $\partial a_j(z)/\partial z = 0$. The waveguide number ($N$) of the SSH lattice is set to be 17 according to the actual experimental design. The motivation for choosing an odd number of $N$ is to avoid the formation of TES at sites $j = \pm N$, which is irrelevant to the manipulation of TZM at the central waveguides. Similar to those used in the simulation in Fig. 1, the coupling strengths $K_a$, $K_b$, and $K_c$ are set to be 4.3, 8.1, and 5.7 mm$^{-1}$, which can be satisfied in the sample fabrication. As shown in Fig. 2(a), the c-SSH lattice supports only one TZM mode with a relative propagation constant of strictly zero lying in the bandgap of bulk states. Figure 2(b) shows the normalized amplitude of the TZM and the bulk modes, where the localized TZM mode can be clearly recognized. The TZM has zero overlap with the central site ($j = 0$) while maximum overlap with the sites $j = \pm 1$. Thus, the TZM can be excited through either site $j = +1$ or $j = -1$, or simultaneously through both sites $j = \pm 1$. Even though only one set of TZM exists in our designed system, the initial dynamic phase of the TZM can be different when being excited through sites $j = +1$ or $j = -1$, which is determined by the dynamic phase of the input excitation laser light therein. In case of a simultaneous excitation through both sites $j = \pm 1$, the input laser through each waveguide site ($j = \pm 1$) can separately excite the TZM, in which the dynamic phase of the excited TZM is determined by the corresponding excitation laser light. As such, the interference between the TZMs separately excited through sites $j = \pm 1$ can occur depending on their relative dynamic phase difference. When the dynamic phase difference between the two input sites ($j = \pm 1$) is π, the two separately excited TZMs are in phase, resulting in a constructive interference and thus the existence of TZM in the composite system. In contrast, if the two input light beams are in phase (i.e. the dynamic phase difference is 0), the excited TZMs are out of phase, resulting in destructive interference and thus elimination of the TZM. The excitation efficiency of TZM can be calculated with the input excitation state of $|\psi_{in}\rangle = \frac{1}{\sqrt{2}}(\cdots, e^{i\Delta\varphi}, 0, 1, \cdots)^T$, where $\Delta\varphi$ denotes the dynamic phase difference between the sites $j = \pm 1$ (see Fig. 2(c)). The excitation efficiency increases from 0 to 0.84 when $\Delta\varphi$ varies from 0 to π. Only bulk states can be excited when the phase difference is $\Delta\varphi = 0$, as the integral field overlap between the input state with a symmetric profile and the TZM with an antisymmetric profile is 0. When $\Delta\varphi$ reaches π, the maximum efficiency is still smaller than 1, which is attributed to the fact that the field of the TZM is not fully confined within the waveguide sites $j = \pm 1$, but slightly extends to the bulk waveguide array region.

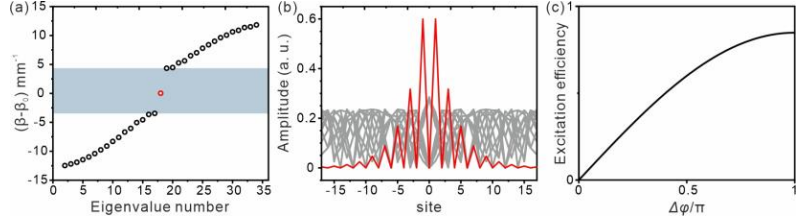

Fig. 2. (a) Calculated spectrum of the designed c-SSH waveguide array showing a TZM located within the bandgap. (b) Amplitude distribution of the normalized wavefunction ($\langle\psi_i|\psi_i\rangle = 1$) of TZM (blue line) and bulk states (gray lines) in the waveguide array. (c) Excitation efficiency of the TZM versus the dynamic phase difference of the input light through waveguides $j = \pm 1$.

Silicon nitride (SiN$_x$) waveguide arrays were fabricated to examine the efficiency of the dynamic phase modulated TZM excitation, as shown in Fig. 3. The width, height and gaps of the waveguide array are consistent with the simulations in Fig. 1(b). The lengths of all waveguides are 500 μm. The waveguide arrays were integrated with several on-chip functional components including multimode interference (MMI) couplers, unbalanced injection arms and a waveguide taper. Bulk and nontrivial topological states can be selectively excited and switched simply by changing the dynamic phase difference of the incident light. In brief, a tapered waveguide was designed to couple laser light to a MMI coupler through a single-mode waveguide, by which the laser light can be split into two unbalanced arms, as is schematically shown in Fig. 3(a). The phase difference generated in the unbalanced arms can be expressed as $\Delta\varphi = n_{eff}k_0\Delta L = 2\pi\lambda^{-1}n_{eff}\Delta L$, where $n_{eff}$ is the effective refractive index, $\Delta L$ is the length difference of the unbalanced arms, and $k_0$ ($2\pi/\lambda$) is the propagation constant in free space. The length difference $\Delta L = L_1 - L_2$ of the two unbalanced arms was designed to be 0, 10, and 30 μm, respectively, to demonstrate the impact of dynamic phase difference on the excitation of the TZM. Based on the simulation in Fig. 2(b), the TZM can be expressed as $\langle\psi_{TZM}| = (\cdots, -0.6, 0, 0.6, \cdots)$. The excitation efficiency is defined as

$$\eta = \langle\psi_{TZM}|\psi_{ini}\rangle = 0.42(1 - cos(\Delta\varphi)). \tag{6}$$

One can see that the coupling efficiency is a function of $\Delta\varphi$ which is determined by both the input laser wavelength $\lambda$ and the $\Delta L$.

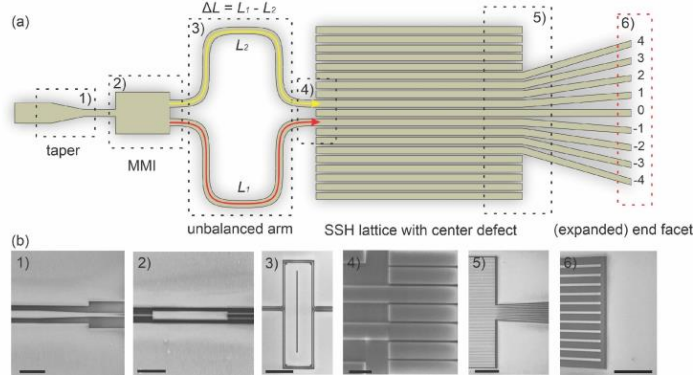

Fig. 3. (a) Schematic diagram of our designed sample and corresponding (b) SEM images of 1) tapered input waveguide, 2) 1×2 MMI coupler, 3) unbalanced arms, 4) injection region to the waveguide array, 5) end facet of the waveguide array, and 6) expanded end facet. Corresponding scale bars from 1) to 6) are 10, 20, 60, 2, 30 and 20 μm.

The effect of the dynamic phase difference of the input laser light on the mode field distributions was simulated using the beam propagation method [36], as shown in Fig. 4. The calculated mode field distribution from site 0 to ±5 was shown as a function of $\Delta\varphi/2\pi$. Consistent with the CMT analysis, in the cases of $\Delta\varphi = 0$ and $\pi$, the input laser beams preferentially excite either the bulk and the TZM modes, respectively. In contrast to the case of $\Delta\varphi = 0$, the mode field excited by input laser beams with $\Delta\varphi = \pi$ is strongly localized at the waveguides $j = \pm 1$, indicating the effective excitation of the TZM. As depicted in Fig. 4(a), when $\Delta\varphi = \pi/2$, a nearly uniform distribution of mode field was observed at the end facets, representing a mixed state of TZM and bulk modes. Intriguingly, considering the current design of the structure, constructive interference of TZMs occurs in the case of out-of-phase of the two input laser beams, while destructive interference happens in the case of in-phase of the input laser beams. The constructive and destructive interference result in effective excitation of the TZM and bulk states, respectively.

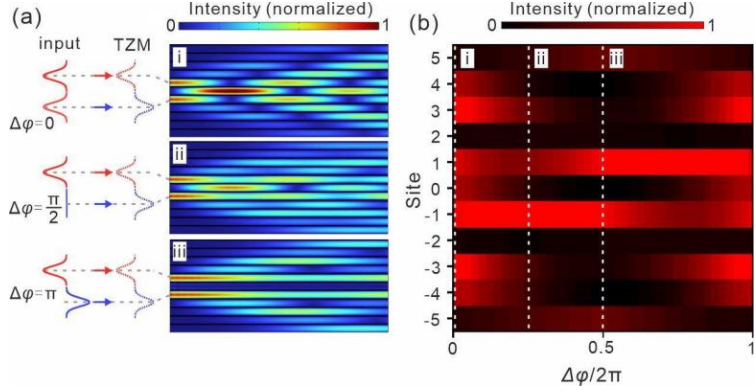

Fig. 4. (a) Schematic diagrams depict the instantaneous phase difference ($\Delta\varphi = 0$, $\pi/2$ and $\pi$) between the two input light beams compared with that of TZM at the input waveguides $j = \pm 1$. The corresponding mode distributions (i-iii) in the waveguide array show a maximum excitation of the TZM at $\Delta\varphi = \pi$. (b) The mode distributions at the end facet of the waveguide array as a function of $\Delta\varphi/2\pi$. The mode distribution cases (i-iii) are marked in dashed lines showing the transition from bulk to non-trivial topological mode.

Figure 5 displays the systematic experimental results of the mode field distributions at the end facets of the waveguide arrays by varying $\Delta L$ and $\lambda$. In the case of $\Delta L = 0$ μm, the excitation efficiency of the TZM is zero according to Eq. (7) and the mode field shows a constant excitation of bulk modes in the entire spectral range of 780 to 800 nm. When there is a path length difference, for instance, $\Delta L = 10$ μm, a dynamic phase difference occurs after propagating along the unbalanced arms, which is also dependent on wavelength $\lambda$. As shown in Fig. 5(b), a significant difference in the field distribution is found at $\lambda_{b1}$ (dynamic phase difference $\Delta\varphi = \pi$) and $\lambda_{b2}$ (dynamic phase difference $\Delta\varphi = 2\pi$, i.e. 0), corresponding to the excitation of the TZM and of bulk modes, respectively. The wavelength-dependent dynamic phase difference generated in the unbalanced light paths can be further verified using a larger length difference $\Delta L = 30$ μm, where more periods of $\Delta\varphi = 0 - \pi$ (i.e. $\Delta\varphi/2\pi = 0 - 0.5$) can occur in the same spectral range, as illustrated in Fig. 5(c). As such, more periods of evolution from TZM to bulk states are observed as marked from $\lambda_{c1}$ to $\lambda_{c3}$. The experimental results agree well with the simulation results shown in the top panel of Fig. 5(c). These results clearly demonstrate the manipulation of topological modes by changing the dynamic phase difference, which is highly promising for practical applications. For instance, by designing multiple unbalanced arms connected to the waveguides $j = \pm 1$, topological or bulk mode at the same wavelength can be selectively excited in a single device by launching the incident light into certain arm pairs. Alternatively, the topological modes can also be controlled by applying external magnetic field when the unbalanced arm waveguides are made of responsive materials [37, 38], which would greatly extend their applications for controlling and utilizing the topological states in on-chip photonic devices.

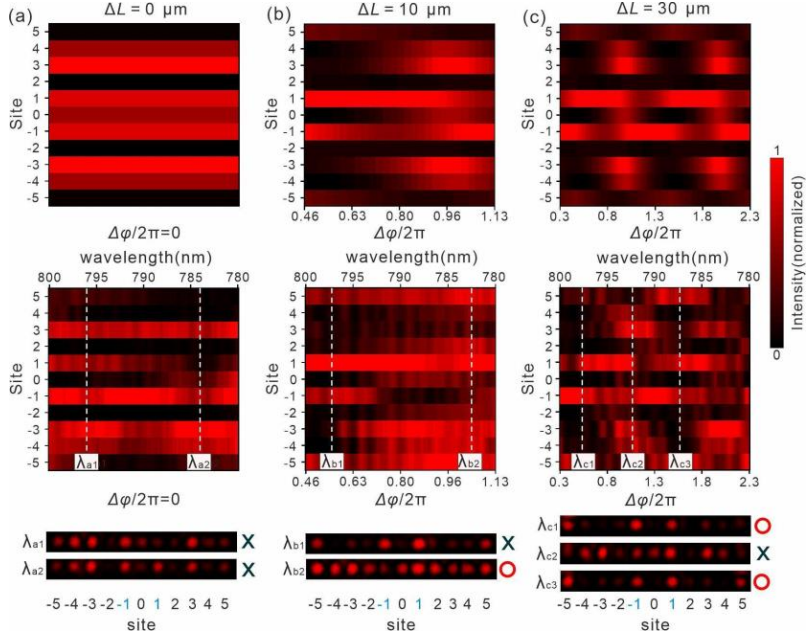

Fig. 5. Simulation (top panel) and experimental (middle panel) results of the mode field distributions in waveguide sites (0 to ±5) versus the phase difference $\Delta\varphi/2\pi$ for the cases of (a) $\Delta L = 0$ μm, (b) 10 μm, and (c) 30 μm. Bottom panel: mode distributions recorded from the waveguide end facets show typical bulk (labeled with black crosses) and TZM (labeled with red circles) modes, corresponding to the cases marked by dashed lines in the middle panels.

## 3. CONCLUSION

In summary, we have proposed and demonstrated the tunable excitation of a topological zero mode with non-trivial geometric phase, which is enabled by changing the dynamic phase difference in composite topological waveguide arrays. With this approach, we show that bulk or topological modes can be selectively excited by changing the dynamic phase differences which results in either a destructive or constructive interference in the designed composite system. Both experimental and theoretical investigations have been carried out, giving consistent results. To the best of our knowledge, the strategy proposed here is currently one of the simplest methods for the controllable excitation of topological state in photonic lattices, which builds a bridge between the geometric phase and dynamic phase of light. The present work can also inspire the research of controlling the mode excitation in non-Hermitian topological systems [39], where the light flow direction can be switched by manipulating the dynamic phase. This research provides a new tuning knob for the manipulation of topological states in composite photonic devices, holding great potential for broad applications where the dynamic phase, geometric phase and wavelength play crucial roles in regulating light flow in on-chip topological devices.

## 4. Experimental Section

The composite waveguide arrays were fabricated by electron beam lithography (Voyager, Raith, GmbH) followed by reactive-ion etching (Plasma Lab 100, Oxford Instruments PLC). In brief, a 220 nm thick $SiN_x$ layer was firstly deposited on $Si/SiO_2$ wafer by plasma-enhanced chemical vapor deposition (ICP: 1200 W; $SiH_4$ [5% in He]: 62 sccm; $N_2$: 20 sccm; Ar: 35 sccm; temperature: 275 °C; time: 1650 s; ICPECVD SI 500 D). A positive resist (SML 300, EM resist LTD, Macclesfield, UK) was spun onto the samples (1700 rpm@60 s), which were exposed by 50 kV electron beam in an EBL system (Voyager, Raith, GmbH, Dortmund, Germany) with a dose of 800 μC $cm^{-2}$ A 30 min post-bake at 80°C in a convection oven and a 30 s $O_2$ treatment were implanted to smooth the edge of the structure and to remove residual photoresist on the exposed surfaces. The waveguides were then etched in an ICP-RIE (Plasma Lab 100, Oxford Instruments plc, Abingdon, UK) in RIE mode, with Ar: 20 sccm, $CHF_3$: 3 sccm, $CF_4$: 10 sccm, and $O_2$: 4 sccm as reactive gases with other parameters ICP: 0 W, Bias: 175 V, pressure: 0.012 mbar, etching time: 80 s. A layer of PMMA (ARP 679.03, 3000 rpm@60 s) was spun coated on the surface of the waveguides as an index matching as well as a protection layer.

In the measurement setup, a wavelength-tunable Ti:sapphire laser (Equinox, M Squared Lasers, Glasgow, UK) was end-fired into the cleaved facet of the taper waveguide assisted by fiber-port coupler and polarization controller. To image light-scattering patterns from the top view of the waveguide arrays, a 24x high-magnification zoom lens system was used together with an 8 MP monochrome CCD camera (S805MU2, Thorlabs).

**Funding.** Würzburg-Dresden Cluster of Excellence on Complexity and Topology in Quantum Matter−ct.qmat (EXC 2147, project-ID 390858490); German Research Foundation (MA 7968/2-1); Science and Technology Innovation Commission of Shenzhen (JCYJ20220531095604009, RCYX20221008092907027); Fundamental Research Funds for the Central Universities (HIT.OCEF.2023025).

**Acknowledgments.** The authors thank M. Bauer, L. Raith and R. Engelhard for the technical support.

**Data availability.** Data underlying the results presented in this paper are not publicly available at this time but may be obtained from the authors upon reasonable request.

## References

1. M. V. Berry, "Quantal phase factors accompanying adiabatic changes," Proceedings of the Royal Society of London. A. Mathematical and Physical Sciences **392**, 45-57 (1984).
2. J. J. N. Anandan, "The geometric phase," Nature **360**, 307-313 (1992).
3. K. v. Klitzing, G. Dorda, and M. Pepper, "New method for high-accuracy determination of the fine-structure constant based on quantized Hall resistance," Physical review letters **45**, 494 (1980).
4. O. Hosten, and P. Kwiat, "Observation of the spin Hall effect of light via weak measurements," Science **319**, 787-790 (2008).
5. K. Y. Bliokh, D. Smirnova, and F. Nori, "Quantum spin Hall effect of light," Science **348**, 1448-1451 (2015).
6. D. C. Tsui, H. L. Stormer, and A. C. Gossard, "Two-dimensional magnetotransport in the extreme quantum limit," Physical Review Letters **48**, 1559 (1982).
7. C. A. Rosiek, G. Arregui, A. Vladimirova et al., "Observation of strong backscattering in valley-Hall photonic topological interface modes," Nature Photonics, 1-7 (2023).

8. Z. Wang, Y. Chong, J. D. Joannopoulos et al., "Observation of unidirectional backscattering-immune topological electromagnetic states," Nature **461**, 772-775 (2009).
9. M. C. Rechtsman, "Reciprocal topological photonic crystals allow backscattering," Nature Photonics **17**, 383-384 (2023).
10. P. Roushan, J. Seo, C. V. Parker et al., "Topological surface states protected from backscattering by chiral spin texture," Nature **460**, 1106-1109 (2009).
11. Y. Zeng, U. Chattopadhyay, B. Zhu et al., "Electrically pumped topological laser with valley edge modes," Nature **578**, 246-250 (2020).
12. M. A. Bandres, S. Wittek, G. Harari et al., "Topological insulator laser: Experiments," Science **359**, eaar4005 (2018).
13. L. Yang, G. Li, X. Gao et al., "Topological-cavity surface-emitting laser," Nature Photonics **16**, 279-283 (2022).
14. Y. Kang, X. Ni, X. Cheng et al., "Pseudo-spin–valley coupled edge states in a photonic topological insulator," Nature communications **9**, 3029 (2018).
15. O. Bleu, D. Solnyshkov, and G. Malpuech, "Quantum valley Hall effect and perfect valley filter based on photonic analogs of transitional metal dichalcogenides," Physical Review B **95**, 235431 (2017).
16. J.-X. Fu, J. Lian, R.-J. Liu et al., "Unidirectional channel-drop filter by one-way gyromagnetic photonic crystal waveguides," Applied Physics Letters **98** (2011).
17. G.-J. Tang, X.-D. Chen, F.-L. Shi et al., "Frequency range dependent topological phases and photonic detouring in valley photonic crystals," Physical Review B **102**, 174202 (2020).
18. S. A. Skirlo, L. Lu, and M. Soljačić, "Multimode one-way waveguides of large Chern numbers," Physical review letters **113**, 113904 (2014).
19. T. Ma, and G. Shvets, "Scattering-free edge states between heterogeneous photonic topological insulators," Physical Review B **95**, 165102 (2017).
20. R. Bach, D. Pope, S.-H. Liou et al., "Controlled double-slit electron diffraction," New Journal of Physics **15**, 033018 (2013).
21. M. Born, and E. Wolf, *Principles of optics: electromagnetic theory of propagation, interference and diffraction of light* (Elsevier, 2013).
22. J.-L. Tambasco, G. Corrielli, R. J. Chapman et al., "Quantum interference of topological states of light," Science advances **4**, eaat3187 (2018).
23. A. Blanco-Redondo, B. Bell, D. Oren et al., "Topological protection of biphoton states," Science **362**, 568-571 (2018).
24. S. Mittal, V. V. Orre, E. A. Goldschmidt et al., "Tunable quantum interference using a topological source of indistinguishable photon pairs," Nature Photonics **15**, 542-548 (2021).
25. A. Blanco-Redondo, I. Andonegui, M. J. Collins et al., "Topological optical waveguiding in silicon and the transition between topological and trivial defect states," Physical review letters **116**, 163901 (2016).
26. Q. Cheng, Y. Pan, Q. Wang et al., "Topologically protected interface mode in plasmonic waveguide arrays," Laser & Photonics Reviews **9**, 392-398 (2015).
27. J. Wang, S. Xia, R. Wang et al., "Topologically tuned terahertz confinement in a nonlinear photonic chip," Light: Science & Applications **11**, 152 (2022).
28. T. Chen, Y. Yu, Y. Song et al., "Distinguishing the topological zero mode and Tamm mode in a microwave waveguide array," Annalen der Physik **531**, 1900347 (2019).
29. S. Suntsov, K. Makris, D. Christodoulides et al., "Observation of discrete surface solitons," Physical review letters **96**, 063901 (2006).
30. T. Goto, A. Dorofeenko, A. Merzlikin et al., "Optical Tamm states in one-dimensional magnetophotonic structures," Physical review letters **101**, 113902 (2008).
31. R. G. Hunsperger, *Integrated optics: theory and technology* (Springer Science & Business Media, 2009).
32. P. St-Jean, V. Goblot, E. Galopin et al., "Lasing in topological edge states of a one-dimensional lattice," Nature Photonics **11**, 651-656 (2017).
33. P. Gagel, O. A. Egorov, F. Dzimira et al., "An electrically pumped topological polariton laser," arXiv preprint arXiv:2402.00639 (2024).
34. P. Delplace, D. Ullmo, and G. Montambaux, "Zak phase and the existence of edge states in graphene," Physical Review B **84**, 195452 (2011).
35. M. Tang, J. Wang, S. Valligatla et al., "Symmetry-Induced Selective Excitation of Topological States in Su–Schrieffer–Heeger Waveguide Arrays," Advanced Photonics Research **4**, 2300113 (2023).
36. K. Okamoto, *Fundamentals of optical waveguides* (Elsevier, 2021).
37. D. M. Wu, M. L. Solomon, G. V. Naik et al., "Chemically responsive elastomers exhibiting unity-order refractive index modulation," Advanced Materials **30**, 1703912 (2018).
38. M. Wang, and Y. Yin, "Magnetically responsive nanostructures with tunable optical properties," Journal of the American Chemical Society **138**, 6315-6323 (2016).
39. S. Weidemann, M. Kremer, T. Helbig et al., "Topological funneling of light," Science **368**, 311-314 (2020).